\RequirePackage{lineno}
\documentclass[twocolumn,nofootinbib,amsmath,amssymb,prc,aps,superscriptaddress,showpacs]{revtex4}
\usepackage[]{inputenc}
\usepackage[dvips]{graphicx}
\usepackage{color}
\usepackage{bm}
\usepackage{mathpazo}

\begin{document}

\title{Predictions for isobaric collisions at $\sqrt{s_{_{\rm NN}}}$ = 200 GeV from a multiphase transport model}

\author{Wei-Tian Deng}
\affiliation{School of Physics, Huazhong University of Science and Technology, Wuhan 430074, China}
\author{Xu-Guang Huang}
\affiliation{Physics Department and Center for Particle Physics and Field Theory, Fudan University, Shanghai 200433, China}
\affiliation{Key Laboratory of Nuclear Physics and Ion-beam Application (MOE), Fudan University, Shanghai 200433, China}
\author{Guo-Liang Ma}
\affiliation{Shanghai Institute of Applied Physics, Chinese Academy of Sciences, Shanghai 201800, China}
\author{Gang Wang}
\affiliation{Department of Physics and Astronomy, University of California, Los Angeles, California 90095, USA}


\begin{abstract}

The isobaric collisions of $^{96}_{44}$Ru + $^{96}_{44}$Ru and $^{96}_{40}$Zr + $^{96}_{40}$Zr have recently been proposed to discern the charge separation signal of the chiral magnetic effect (CME). In this article, we employ the string melting version of a multiphase transport model to predict various charged-particle observables, including $dN/d\eta$, $p_T$ spectra, elliptic flow ($v_2$), and particularly possible CME signals in Ru + Ru and Zr + Zr collisions at $\sqrt{s_{_{\rm NN}}}$ = 200 GeV. Two sets of the nuclear structure parametrization have been explored, and the difference between the two isobaric collisions appears to be small, in terms of $dN/d\eta$, $p_T$ spectra, and $v_2$ for charged particles. We mimic the CME by introducing an initial charge separation that is proportional to the magnetic field produced in the collision, and study how the final-state interactions affect the CME observables. The relative difference in the CME signal between the two isobaric collisions is found to be robust, insensitive to the final-state interactions.

\end{abstract}

\maketitle

\section{I. Introduction}
\label{sec:intro}
In the initial stage of a non-central high-energy heavy-ion collision, the spectator protons in the two beams pass by each other at very high speeds, creating in the overlap region a strong magnetic field with the order of magnitude of $eB \sim m_{\pi}^{2} \sim 10^{18}G$~\cite{Skokov:2009qp,Voronyuk:2011jd,Deng:2012pc,Deng:2014uja}. If the magnetic field lasts long enough, it will have important impacts on the nuclear matter formed in the successive stages of the collision, e.g., the deconfined quark gluon plasma (QGP)~\cite{Adams:2005dq,Adcox:2004mh}. 
In contrast to the concept of an empty vacuum, the quantum chromodynamics vacua contain fluctuating chromo-electric and chromo-magnetic fields, and instant topological transitions can occur between two neighboring degenerate vacua at high temperatures~\cite{Lee:1974ma,McLerran:1990de}.
Owing to these transitions, metastable domains will emerge in the QGP, with ${\cal P}$ and ${\cal CP}$ locally violated.
In such domains, light quarks manifest a chirality imbalance, characterized by a non-zero chiral chemical potential ($\mu_5$). The chirality imbalance, together with
the strong magnetic field,
will induce an electric current along the direction of the magnetic field, which is called the chiral magnetic effect (CME)~\cite{Kharzeev:2007jp,Fukushima:2008xe}; see recent reviews in Refs.~\cite{Kharzeev:2015znc,Wang:2016mkm,Huang:2015oca,Hattori:2016emy}.

On average, the magnetic field direction is perpendicular to the reaction plane (defined by the impact parameter and the beam momenta), so the CME will form a charge separation with respect to the reaction plane.
The searches for the CME-related charge separation have been performed in experiments at both the BNL Relativistic Heavy Ion Collider (RHIC)~\cite{Abelev:2009ac,Abelev:2009ad,Adamczyk:2014mzf} and the CERN Large Hadron Collider (LHC)~\cite{Abelev:2012pa,Acharya:2017fau}.
A main-stream experimental observable is the charge-dependent azimuthal correlator~\cite{Voloshin:2004vk}, $\gamma_{\alpha\beta}=\langle \cos(\phi_{\alpha}+\phi_{\beta}-2\Psi_{\rm RP})\rangle$, where $\phi_{\alpha(\beta)}$ is the azimuthal angle of a charged particle $\alpha(\beta)$, and $\Psi_{\rm RP}$ is the angle of the reaction plane. Although these measurements exhibit features that qualitatively meet the CME expectation, there exist ambiguities in the interpretation of the experimental data, because elliptic-flow driven backgrounds also contribute to the measured correlator~\cite{Bzdak:2010fd,Liao:2010nv,Schlichting:2010qia,Wang:2009kd}. 

Among different schemes proposed to discern the true CME signal, the isobaric collisions are favored, e.g., $^{96}_{44}$Ru + $^{96}_{44}$Ru and $^{96}_{40}$Zr + $^{96}_{40}$Zr at the same beam energy. Ru and Zr nuclei have a $10\%$ difference in the number of protons, but the same number of nucleons, which will roughly lead to a $20\%$ difference in the CME signal in $\gamma_{\alpha\beta}$,  and will keep the elliptic-flow induced backgrounds similar~\cite{Voloshin:2010ut,Deng:2016knn,Skokov:2016yrj,Huang:2017azw}; see also a recent study using the anomalous hydrodynamics~\cite{Shi:2017cpu} and the discussion about the role played by the isobaric density distributions in the CME search~\cite{Xu:2017zcn}. Our recent work demonstrates that isobaric collisions can provide an ideal tool to disentangle the CME signal from the flow backgrounds~\cite{Deng:2016knn}, if the final-state interactions are ignored when estimating the CME observables. However, a previous study using a multi-phase transport (AMPT) model shows that the final-state interactions, including parton cascade, hadronization and resonance decays, can significantly suppress the initial charge separation in Au + Au collisions~\cite{Ma:2011uma}. Therefore, it is important to run a full numerical simulation to take into account all the above-mentioned final-state interactions. 

In this work, we simulate Ru + Ru and Zr + Zr collisions at $\sqrt{s_{_{\rm NN}}}$ = 200 GeV with the AMPT model, and introduce an initial charge separation to mimic the CME. We will make predictions for some basic observables, e.g., multiplicity, $p_T$ spectra, and elliptic flow for charged particles, as well as for the CME-related observables. 
This paper is organized as follows. In Sec.~\ref{sec:setup} we introduce the general setup of the simulation. The numerical results and discussions are presented in Sec.~\ref{sec:results}, and the summary is in Sec.~\ref{sec:summary}.

\section{General Setup}
\label{sec:setup}
\subsection{The AMPT model}
\label{sec:model}
The AMPT model used in this work is implemented with the string melting mechanism~\cite{Lin:2004en}. AMPT covers four main stages of high-energy heavy-ion collisions: the initial condition, parton cascade, hadronization, and hadronic rescatterings. The initial condition is obtained from the HIJING model~\cite{Wang:1991hta,Gyulassy:1994ew}, which includes the spatial and momentum distributions of minijet partons and soft string excitations. The minijets and soft string excitations fragment into hadrons according to the Lund string fragmentation~\cite{Sjostrand:2000wi}. Then all hadrons are converted to quarks according to the flavor and spin structures of their valence quarks, leading to a formation of a quark and anti-quark plasma. Next, the parton evolution is simulated by Zhang's parton cascade (ZPC) model~\cite{Zhang:1997ej}, where the partonic cross section is controlled by the strong coupling constant and the Debye screening mass. AMPT  recombines partons via a simple coalescence model to produce hadrons when partons freeze-out. The dynamics of subsequent hadronic rescatterings is then described by the ART model~\cite{Li:1995pra}. 

In a recent work by Ma and Lin~\cite{Ma:2016fve}, the string melting version of AMPT takes a universal setting of tuned parameters, and achieves a reasonably good reproduction of $dN/d\eta$, $p_{\rm T}$-spectra, azimuthal anisotropies, and factorization ratios for longitudinal correlations in A+A collisions at both RHIC and the LHC energies. However, the current ART model does not conserve electric charge. Therefore we turn off the hadron evolution (but keep resonance decays) to ensure charge conservation, which is important for the study of the CME-related observables.

\subsection{Description of $^{96}_{44}$Ru and $^{96}_{40}$Zr}

The spatial distribution of nucleons in the rest frame of either $^{96}_{44}$Ru or $^{96}_{40}$Zr is described with the following Woods-Saxon form (in spherical coordinates)~\cite{Deng:2016knn},
\begin{eqnarray}
\rho(r,\theta)=\frac{\rho_0}{1+\exp{[(r-R_0-\beta_2 R_0 Y^0_2(\theta))/a]}},
\end{eqnarray}
where $\rho_0=0.16$ fm$^{-3}$ is the normal nuclear density, $R_0$ and $a$ denote the ``radius" of the nucleus and
the surface diffuseness parameter, respectively, and $\beta_2$ represents the deformity of the nucleus. The parameter $a$ is almost identical
for Ru and Zr: $a\approx 0.46$ fm. However, the $\beta_2$ values for Ru and Zr are unclear within the current knowledge. There are two available sources of $\beta_2$: e-A scattering experiments~\cite{Raman:1201zz,Pritychenko:2013gwa} and comprehensive model
deductions~\cite{Moller:1993ed}. According to the former (which will be referred to as set 1), Ru is more deformed
($\beta_2^{\rm Ru} = 0.158$) than Zr ($\beta_2^{\rm Zr} = 0.08$); whereas the latter (set 2) argues that $\beta_2^{\rm Ru}=0.053$ is smaller than $\beta_2^{\rm Zr}=0.217$,  which is opposite to set 1. Throughout the following studies about the basic observables (see Sec.~\ref{basic}), we will show the consequence of this uncertainty.  For the CME-related studies (see Sec.~\ref{cmeresult}), we will only present the results with $\beta_2$ values from set 2, whereby the larger deformity difference between Ru and Zr makes more visible effects on the elliptic-flow-induced backgrounds.

\subsection{Modeling the initial charge separation}
\label{sect:cs}
The AMPT model itself has no mechanisms to generate the CME. Thus we have to artificially introduce a charge separation into the initial state. To separate a given fraction of the initial charges, we adopt a global charge separation scheme, which was first employed in Ma and Zhang's previous work for Au + Au collisions~\cite{Ma:2011uma}. Note that the results from a local charge separation scheme have been shown to be consistent with those from the global one~\cite{Shou:2014zsa}. We simulate a CME-like initial charge separation by switching the $p_y$ values of a fraction of the downward moving $u$ quarks with those of the upward moving $\bar{u}$ quarks, and likewise for $\bar{d}$ and $d$ quarks. 
In this way, the total momentum is conserved.
The coordinate
system is such that ``upward" and ``downward" are with respect to the
$y$-axis, which is perpendicular to the reaction plane.
The fraction $f$ is defined as,
\begin{equation}
f = \frac{N_{\uparrow(\downarrow)}^{+(-)}-N_{\downarrow(\uparrow)}^{+(-)}}{N_{\uparrow(\downarrow)}^{+(-)}+N_{\downarrow(\uparrow)}^{+(-)}},
 \label{eq-f}
\end{equation}
where $N$ is the number of particles of a given species, $+$ and $-$ denote positive and negative charges, respectively, and $\uparrow$ and $\downarrow$ represent the moving directions. 

As the CME-induced current is proportional to $\mu_5 B_y$, where $B_y$ is the $y$-component of the magnetic field, the separated charge $\Delta Q$ is expected to be $\propto \mu_5 B_y \tau_B S$, with $\tau_B$ the life time of the magnetic field and $S\propto A^{2/3}$ the area of  the overlap region, where $A$ is the nucleon number of the colliding nucleus. We further estimate that $f\sim \Delta Q/Q\propto \mu_5 B_y A^{-1/3}$ for a finite $\mu_5$, which should also depend on $A$. If the local ${\cal P}$-odd domains are generated with their $\mu_5$ bearing random signs, the global $\mu_5$ will be proportional to $A^{-1/2}$. In consideration of possible quenching effects, $\mu_5$ would be further suppressed. Our simulation shows that $\mu_5\propto A^{-1}$ could well fit the experimental data from Au + Au and Cu + Cu collisions, in which case, $f\propto B_y A^{-4/3} $. Since $B_y$ is a function of impact parameter ($b$), the fraction $f$ should depend on $b$ or centrality of the collision.

\begin{figure}
\includegraphics[scale=0.45]{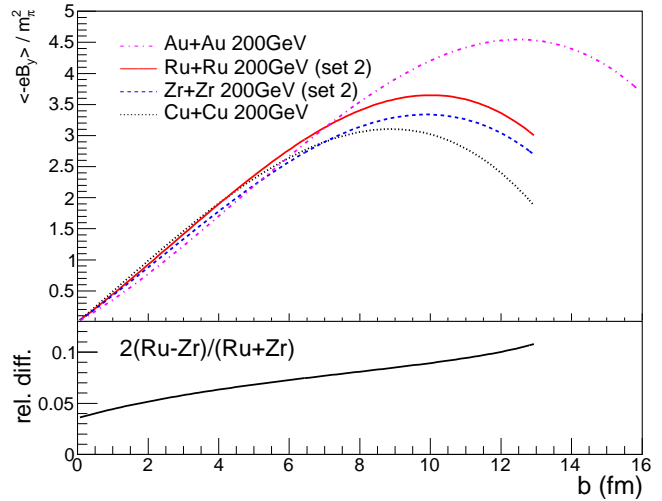}
\caption{Upper: the impact parameter dependence of the magnetic field in different heavy-ion collisions at $\sqrt{s_{_{\rm NN}}}$ = 200 GeV. Lower: the relative difference in the magnetic field between Ru + Ru and Zr + Zr collisions.}
 \label{fig-Bfields}
\end{figure}

The initial magnetic field can be calculated with the Lienard-Wiechert potential~\cite{Deng:2012pc}. The upper panel of Fig.~\ref{fig-Bfields} depicts the impact parameter dependence of the event-averaged magnetic field at the center of the overlap region, $\langle B_y\rangle$, for Au+Au, Cu+Cu, Ru + Ru, and Zr + Zr collisions at $\sqrt{s_{_{\rm NN}}}$ = 200 GeV. The parameters from set 2 are adopted for Ru + Ru and Zr + Zr. Note that the corresponding results using set 1 are quite similar to that using set 2~\cite{Deng:2016knn}. The relative difference~\footnote{The relative difference in a quantity $X$ between Ru + Ru and Zr + Zr collisions is defined by $R_X \equiv 2(X^{\rm Ru+Ru}-X^{\rm Zr+Zr})/(X^{\rm Ru+Ru}+X^{\rm Zr+Zr})$.} in $\langle B_y\rangle$ between Ru + Ru and Zr + Zr is presented in the lower panel of Fig.~\ref{fig-Bfields}. At large $b$, $\langle B_y\rangle$ in Ru + Ru collisions is larger than that in Zr + Zr collisions by roughly $10\%$. This is expectedly consistent with the fact that a Ru nucleus contains $10\%$ more protons than of a Zr nucleus.

\begin{figure}
\includegraphics[scale=0.45]{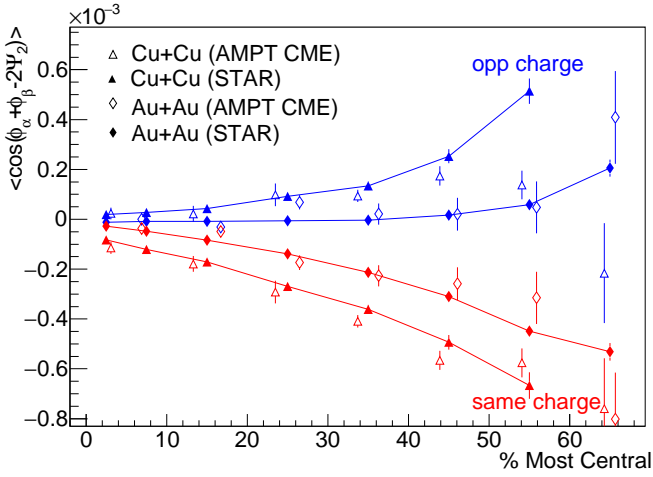}
\caption{The centrality dependence of $\gamma_{\alpha\beta}=\langle \cos(\phi_{\alpha}+\phi_{\beta}-2\Psi_{2})\rangle$ for Au + Au and Cu + Cu collisions at $\sqrt{s_{_{\rm NN}}}$ = 200 GeV, where $\Psi_{2}$ is the second-order event plane. The open symbols represent the results from the AMPT model with a given initial charge separation, and the solid symbols represent the STAR data~\cite{Abelev:2009ac}. }
 \label{fig-PVfit}
\end{figure}

\begin{figure}
\includegraphics[scale=0.45]{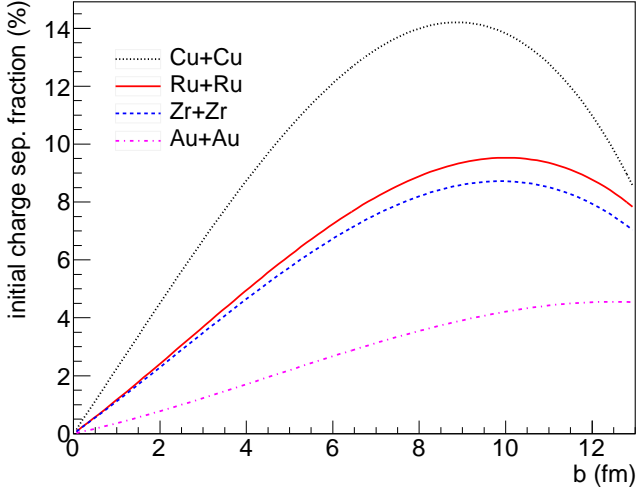}
\caption{The impact parameter dependence of the initial charge separation fraction for four different collision systems at $\sqrt{s_{_{\rm NN}}}$ = 200 GeV, as discussed in Sec.~\ref{sect:cs}.}
 \label{fig-percentbfm}
\end{figure}

Next, we use the AMPT model to fit the Au + Au and Cu + Cu data of the charge dependent correlator $\gamma_{\alpha\beta}=\langle \cos{(\phi_{\alpha}+\phi_{\beta}-2\Psi_{2})}\rangle$ from the STAR Collaboration~\cite{Abelev:2009ac,Abelev:2009ad}, by assuming that the initial charge separation fraction obeys $f\propto B_y A^{-4/3}$. We calculate the angle of event plane $\Psi_2$ by
\begin{eqnarray}
\Psi_{2}&=&\frac{\mathrm{arctan2}(\left\langle r^{2}\sin(2\varphi) \right\rangle, \left\langle r^{2}\cos(2\varphi) \right\rangle)+\pi}{2}
\label{psi2}
\end{eqnarray}%
where $r$ is the displacement of the participating partons from the center of mass, $\varphi$ is the azimuthal angle of the participating partons in the spacial transverse plane, and the bracket means taking average over all initial participating partons~\cite{Alver:2010gr,Ma:2010dv}. The fitting results are shown in Fig.~\ref{fig-PVfit}, which well describe the STAR data, if we choose $f(\%)=1146.1(eB_y/m_\pi^2) A^{-4/3}$ (shown in Fig.~\ref{fig-percentbfm}). This is the form to be further applied to Ru + Ru and Zr + Zr collisions.

\section{Results and Discussions}
\label{sec:results}
In this section, we present the numerical results from the AMPT simulations of the Ru + Ru and Zr + Zr collisions at $\sqrt{s_{_{\rm NN}}}$ = 200 GeV.
Sec.~\ref{basic} shows our predictions for some basic observables including $dN/d\eta$, $p_T$ spectra, and elliptic flow of charged particles. The CME-related observables will be discussed in Sec.~\ref{cmeresult}, including the charge azimuthal correlator $\gamma_{\alpha\beta}$ and the $H$ correlator (see definition below).

\subsection{Basic observables} \label{basic}
\begin{figure}
\includegraphics[scale=0.45]{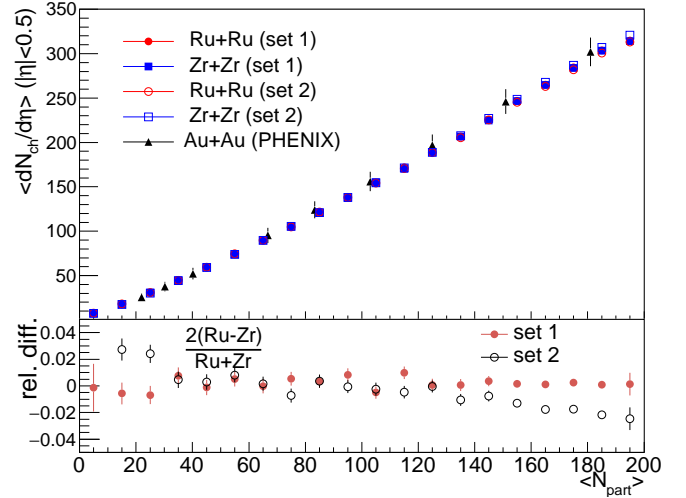}
\caption{Upper: the $\langle N_{\rm part}\rangle$ dependence of the charged particle multiplicity $dN_{\rm ch}/d\eta$ at mid-pseudorapidity  ($|\eta|<0.5$) from the AMPT model with two settings of the isobar collisions at $\sqrt{s_{_{\rm NN}}}$ = 200 GeV. The PHENIX data for Au+Au collisions at $\sqrt{s_{_{\rm NN}}}$ = 200 GeV~\cite{Adler:2004zn} are also shown in comparison. Lower: the relative difference in $dN_{\rm ch}/d\eta$ between Ru + Ru and Zr + Zr.}
 \label{fig-dNdy}
\end{figure}

The upper panel of Fig.~\ref{fig-dNdy} delineates $dN_{\rm ch}/d\eta$ at mid-pseudorapidity  ($|\eta|<0.5$) as a function of the number of participant nucleons, $\langle N_{\rm part}\rangle$, from the string melting mode of AMPT, for two geometry settings of the isobaric collisions at $\sqrt{s_{_{\rm NN}}}$ = 200 GeV. The PHENIX data for Au+Au collisions at $\sqrt{s_{_{\rm NN}}}$ = 200 GeV~\cite{Adler:2004zn} are also shown in comparison. According to our simulations, the particle production yields from the two isobaric systems are close to each other, and they are roughly proportional to $N_{\rm part}$, in the same way as the PHENIX measurements from Au+Au collisions. To better visualize the difference between the two isobaric collisions, the relative difference $R_{{\rm dN/d}\eta}$ is shown in the lower panel of Fig.~\ref{fig-dNdy}. The relative difference is consistent with zero for almost the whole centrality range with set 1, and deviates from zero with set 2. However, the deviation is at most 3\% for set 2, and more severe in the most central and most peripheral collisions, where the geometry deformation seems to play a more prominent role.

\begin{figure}
\includegraphics[scale=0.45]{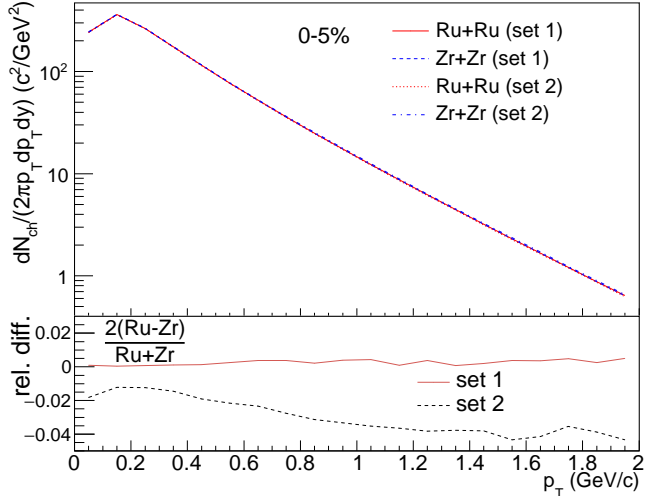}
\caption{Upper: the $p_{T}$ spectra of the charged hadrons at mid-pseudorapidity  ($|\eta|<1$) from the string melting model of AMPT with two settings of central ($0-5\%$) isobar collisions at $\sqrt{s_{_{\rm NN}}}$ = 200 GeV. Lower: the corresponding relative differences.}
 \label{fig-ptspectra}
\end{figure}

The upper panel of Fig.~\ref{fig-ptspectra} shows the AMPT results of the $p_{T}$ spectra of charged hadrons at mid-pseudorapidity  ($|\eta|<1$) for two geometry settings of the central ($0-5\%$) isobaric collisions at $\sqrt{s_{_{\rm NN}}}$ = 200 GeV. The $p_{T}$ spectra for Ru + Ru and Zr + Zr are close to each other, and the two settings also display similarity. However, the relative difference demonstrates a slight mismatch between the two collision systems with set 2, as shown in the lower panel in Fig.~\ref{fig-ptspectra}. The deviation of $R_{p_T}$ from zero with set 2 is in line with the relative difference in multiplicity shown in Fig.~\ref{fig-dNdy}. 

\begin{figure}
\includegraphics[scale=0.45]{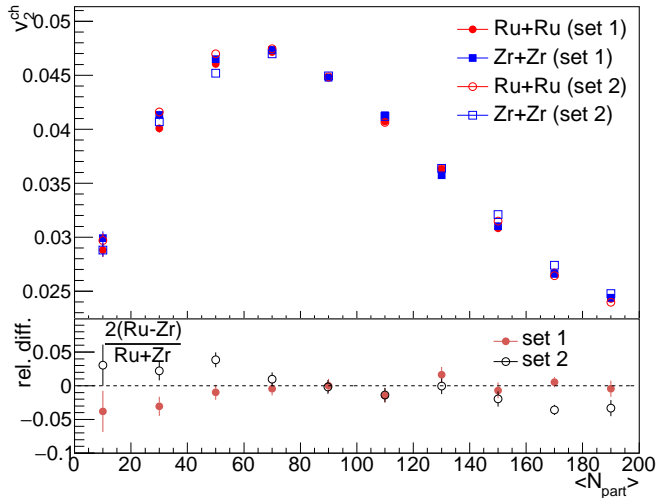}
\caption{Upper: the $N_{\rm part}$ dependence of elliptic flow ($v_2$) of charged hadrons at mid-pseudorapidity  ($|\eta|<1$) from the string melting model of AMPT, with two settings of isobar collisions at $\sqrt{s_{_{\rm NN}}}$ = 200 GeV. Lower: the corresponding relative difference in $v_2$.}
\label{fig-v2cen}
\end{figure}

The upper panel of Fig.~\ref{fig-v2cen} presents the AMPT results for the $\langle N_{\rm part}\rangle$ dependence of elliptic flow ($v_2$) of charged hadrons at mid-pseudorapidity  ($|\eta|<1$) with two geometry settings of isobaric collisions at $\sqrt{s_{_{\rm NN}}}$ = 200 GeV. The event plane method is used to calculate $v_2$, i.e. $v_2=\langle\cos[2(\phi-\Psi_{2})]\rangle$, where $\Psi_{2}$ is obtained from the spatial information of initial partons by Eq. (\ref{psi2}). The lower panel of Fig.~\ref{fig-v2cen} shows the  relative difference in $v_2$ as a function of $N_{\rm part}$, which looks similar to the relative difference in multiplicity, as shown in Fig.~\ref{fig-dNdy}. The geometry deformation seems to affect not only particle yields but also their azimuthal anisotropy.

\subsection{CME-related observables}\label{cmeresult}

This subsection presents our simulation results for the CME-related observables, including the charge dependent correlator $\gamma_{\alpha\beta}=\langle \cos{(\phi_{\alpha}+\phi_{\beta}-2\Psi_{2})}\rangle$ and its CME portion $H_{\alpha\beta}$ (see definition below), from the AMPT model, with the implementation of the initial charge separation. For simplicity, we only show the results with the geometry deformation of set 2.

\begin{figure}
\includegraphics[scale=0.45]{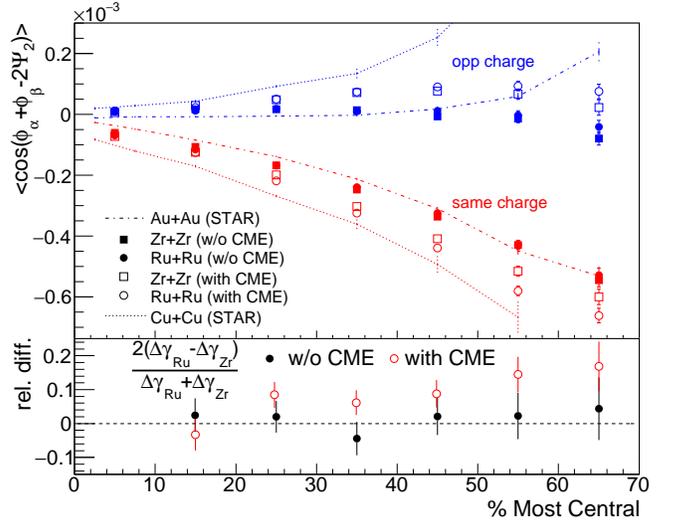}
\caption{Upper: the centrality dependence of the correlator $\gamma_{\alpha\beta}$ for Ru + Ru and Zr + Zr collisions at $\sqrt{s_{_{\rm NN}}}$ = 200 GeV from the AMPT model, with and without introduction of the initial charge separation to mimic the CME. The STAR data for Au + Au and Cu + Cu collisions at $\sqrt{s_{_{\rm NN}}}$ = 200 GeV~\cite{Abelev:2009ac} are also shown with lines in comparison. Lower: the relative difference in $\Delta\gamma$ between the two collisions. Here $\Delta$ means the difference between the opposite-charge and the same-charge correlations.}
 \label{fig-PVRuZr}
\end{figure}

The upper panel of Fig.~\ref{fig-PVRuZr} illustrates the centrality dependence of $\gamma_{\alpha\beta}$ for Ru + Ru and Zr + Zr collisions at $\sqrt{s_{_{\rm NN}}}$ = 200 GeV, from AMPT with and without the imported CME effect. The STAR data for Au+Au and Cu+Cu collisions at $\sqrt{s_{_{\rm NN}}}$ = 200 GeV~\cite{Abelev:2009ac} are also shown in comparison. The event plane angle is determined in the same way as in the previous subsection where $v_2$ is estimated. In the case without the initial charge separation, the Ru + Ru and Zr + Zr collisions generate very similar $\gamma_{\rm opp}$ and $\gamma_{\rm same}$. Once the CME effect is imported, the difference in $\gamma_{\rm opp}$ and $\gamma_{\rm same}$ between the two isobaric systems becomes apparent. Meanwhile, a magnitude ordering of Au + Au $< $ Zr + Zr $<$ Ru + Ru $<$ Cu + Cu is found in the case with the CME effect in the $20-70\%$ centrality range. The lower panel of Fig.~\ref{fig-PVRuZr} shows the relative difference in $\Delta\gamma=\gamma_{\rm opp}-\gamma_{\rm same}$ between Ru + Ru and Zr + Zr collisions. Again, in the case without the initial charge separation, the relative difference is very close to zero; whereas once the initial charge separation is introduced, the relative difference becomes finite with magnitude around $10-20\%$ in the $20-70\%$ centrality range. Our results suggest that the initial charge separation induced by the CME can survive the final-state interactions in Ru + Ru and Zr + Zr collisions, and the relative difference in $\Delta\gamma$ follows a similar trend as that in the projected magnetic field squared~\cite{Bloczynski:2012en,Bloczynski:2013mca,Chatterjee:2014sea,Deng:2016knn}.

\begin{figure}
\includegraphics[scale=0.45]{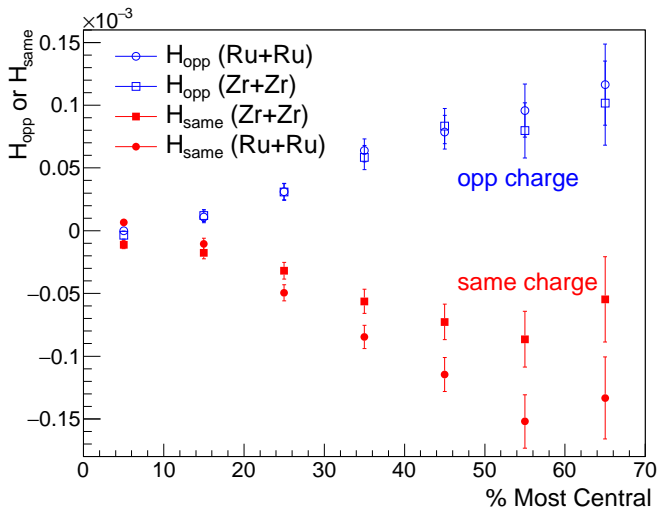}
\caption{The centrality dependence of $H_{\rm opp}$ and $H_{\rm same}$ for Ru + Ru and Zr + Zr collisions at $\sqrt{s_{_{\rm NN}}}$ = 200 GeV.}
 \label{fig-HoppHsame}
\end{figure}

Fig.~\ref{fig-PVRuZr} also reveals that even without the initial CME, the AMPT model can produce a significant amount of $\Delta\gamma_{\alpha\beta}$. This is because AMPT inherently obeys transverse momentum conservation and local charge conservation, which constitute background contributions to $\Delta\gamma_{\alpha\beta}$. To remove the background contributions from the experimental observables, the STAR Collaboration has applied a two-component model to extract the CME contribution from $\gamma_{\alpha\beta}$~\cite{Adamczyk:2014mzf,Bzdak:2012ia}. In our AMPT simulation, the pure CME contribution can be easily extracted by taking a difference between the AMPT results with and without the initial charge separation: $H_{\alpha\beta}=\gamma_{\alpha\beta}^{\rm CME}-\gamma_{\alpha\beta}^{\rm no\, CME}$. Figure~\ref{fig-HoppHsame} presents the AMPT results on the centrality dependence of $H_{\rm opp}$ and $H_{\rm same}$ for Ru + Ru and Zr + Zr collisions at $\sqrt{s_{_{\rm NN}}}$ = 200 GeV. A clear difference in $H$ between the two isobaric systems appears in the $20-70\%$ centrality range.

\begin{figure}
\includegraphics[scale=0.45]{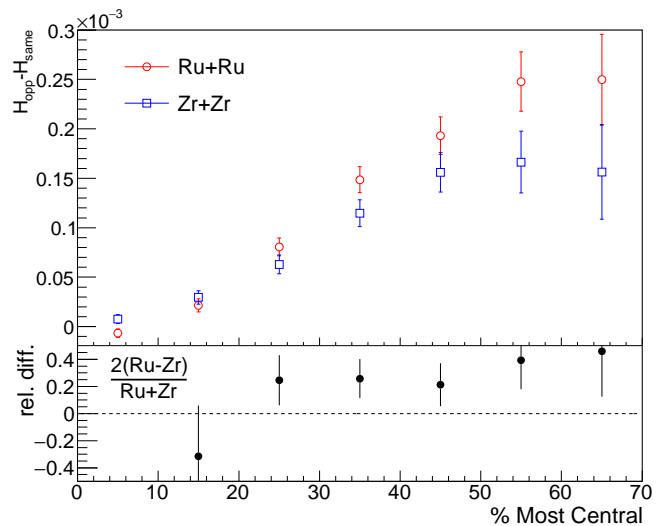}
\caption{Upper: the centrality dependence of $H_{\rm opp}-H_{\rm same}$ for Ru + Ru and Zr + Zr collisions at $\sqrt{s_{_{\rm NN}}}$ = 200 GeV. Lower: the corresponding relative difference.}
 \label{fig-Htotal}
\end{figure}

The difference between the opposite-charge and same-charge correlators, i.e. $\Delta H=\rm H_{\rm opp}-\rm H_{\rm same}$, is shown in the upper panel of Fig.~\ref{fig-Htotal}. $\Delta H$ increases as the collision becomes more peripheral, with the magnitudes in Ru + Ru collisions larger than those in Zr + Zr collisions. The relative difference in $\Delta H$ is presented in the lower panel of Fig.~\ref{fig-Htotal}, showing a similar trend as that in $\Delta\gamma$ in Fig.~\ref{fig-PVRuZr}, but with larger magnitudes. This again indicates that the relative difference in $\Delta\gamma$ can clearly reflect the CME contribution, though the backgrounds strongly dilute the true signal.

\begin{figure}
\includegraphics[scale=0.45]{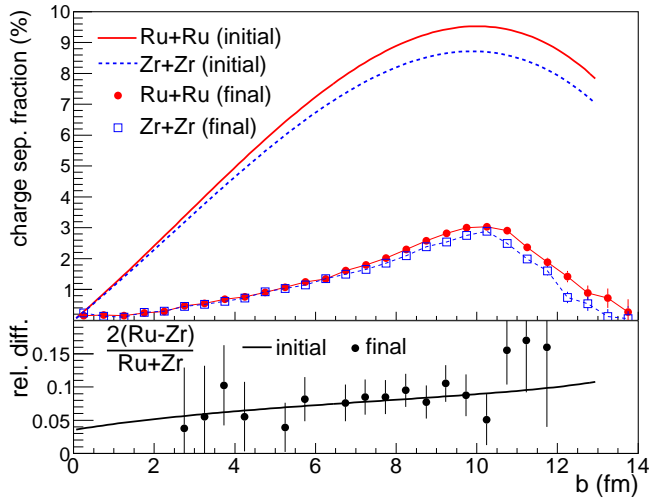}
\caption{Upper: the impact parameter dependence of the final charge separation fraction compared with the initial charge separation fraction for Ru + Ru and Zr + Zr collisions at $\sqrt{s_{_{\rm NN}}}$ = 200 GeV. Lower: the relative difference in the charge separation fraction between Ru + Ru and Zr + Zr collisions for the initial and the final states.}
 \label{fig-finalpercent}
\end{figure}

Since the final-state interactions can significantly reduce the initial CME~\cite{Ma:2011uma}, it is useful to investigate how much of the imported charge separation can survive the final-state interactions. The upper panel of Fig.~\ref{fig-finalpercent} shows the AMPT results on the impact parameter dependence of the initial and final charge separation fractions for Ru + Ru and Zr + Zr collisions. The charge separation fraction is suppressed by the final-state interactions by roughly a factor of 3. However, the final charge separation fraction in Ru + Ru is still higher than that in Zr + Zr. The lower panel of Fig.~\ref{fig-finalpercent} shows the relative difference in the charge separation fraction between Ru + Ru and Zr + Zr collisions, for both the initial and the final states. The relative difference for the final states is consistent with that for the initial states, which means that the final-state interactions can preserve the relative difference between the two collision systems. Therefore, the relative difference is a robust quantity to manifest the CME signal in the isobaric collisions.

\section{Summary}
\label{sec:summary}

Based on a multiphase transport model with an initial charge separation, we present predictions on various observables in two isobaric collisions, $^{96}_{44}$Ru + $^{96}_{44}$Ru and $^{96}_{40}$Zr + $^{96}_{40}$Zr at $\sqrt{s_{_{\rm NN}}}$ = 200 GeV. With two different geometry deformation settings, we find that charged-hadron $dN/d\eta$ scales with $N_{\rm part}$, and weakly depends on the nuclear geometry deformation. The $p_T$ spectra show that the multiplicity difference between Ru + Ru and Zr + Zr collisions is $p_T$-dependent for different geometry deformation settings. Elliptic flow displays a slight difference between the two isobaric systems, which depends on the geometry deformation setting. To mimic the CME, we introduce an initial charge separation in the AMPT model, which is proportional to the magnetic field in each isobaric collision. Although the final-state interactions can reduce the charge separation in each collision, the relative difference in the charge separation between the two isobaric systems is found to be insensitive to the final-state interactions. In particular, the relative difference in $\Delta\gamma$ is nearly zero when there is no initial charge separation, whereas it is finite once an initial charge separation is introduced, with its magnitude between $10-20\%$ for the $20-70\%$ centrality range.
This magnitude is consistent with the previous results in Ref.~\cite{Deng:2016knn}, and supports the proposal that the future isobaric collisions provide an excellent means to disentangle the CME signal from the backgrounds.

{\bf Acknowledgments:} W.-T.D is supported by NSFC with Grant No. 11405066. X.-G.H. is supported by NSFC with Grants No. 11535012 and No. 11675041 and the One Thousand Young Talents Program of China. G.-L.M is supported by NSFC with Grants No. 11522547, No. 11375251, No. 11421505, and the Major State Basic Research Development Program in China with Grant No. 2014CB845404. G.W is supported by the US Department of Energy under Grant No. DE-FG02-88ER40424.

\end{document}